# Photovoltaic response around a unique180° ferroelectric domain wall in single crystalline BiFeO$_3$

C. Blouzon[1], J-Y Chauleau[1], A. Mougin[2], S. Fusil[3] and M. Viret[1]

1) SPEC, CEA, CNRS, Université Paris-Saclay, CEA Saclay, 91191 Gif sur Yvette, France
2) Laboratoire de Physique des Solides, UMR 8502, Bât. 510, 91405 Orsay, France
3) Unité Mixte de Physique, CNRS, Thales, Univ. Paris-Sud, Université Paris-Saclay, 91767 Palaiseau, France

*Abstract*

*Using an experimental setup designed to scan a submicron sized light spot and collect the photogenerated current through larger electrodes, we map the photovoltaic responsein ferroelectric BiFeO$_3$ single crystals. We study the effect produced by a unique 180° ferroelectric domain wall (DW) and show that the photocurrent maps are significantly affected by its presence and shape. The effect is largein itsvicinity and in the Schottky barriers at the interface with the Au electrodes, butno extra photocurrent is observed when the illuminating spot touches the DW, indicating that this particular entity is not the heart of specific photo-electric properties.Using 3D modelling, we argue that the measured effect is due to the spatialdistributionof internal fields which are significantly affected by the charge of the DW due to its distortion.*

Energy harvesting from sunlight is an efficient means of saving fossil fuels which are found in limited amount in the earth's crust. Photovoltaic devices converting light into electricity rely on photo-generation of electron-hole pairs and their subsequent separation and collection. The materials of choice for this are semiconductors with a reasonably small bandgap of the order of 1eV, so that most of the sunlight spectrum can be efficient in kicking electrons from the valence band to the conduction band. An internal electric field,generally realized in a p-n junction, is then required to separate charges and nearby electrodes can collect the current. Ferroelectrics (FE) are thus also natural candidates[1]as they possess the essential necessary ingredients, including the key advantage of an internal built-in electric field.Unfortunately, ferroelectrics have several critical drawbacks including too high bandgaps and recombination rates, which make them uncompetitive for mainstream applications. On a fundamental standpoint, although this field of research started over 30 years ago[2,3,4], the detailedphysical mechanisms driving photocurrents in ferroelectrics are still unclear[5,6,7]. A number of non-centrosymmetric materials exhibit the so calledbulk photovoltaic (BPV) effect whereby light-induced charge carriers are driven by an intrinsic forcelinked to the symmetry of the crystal lattice[3,8] giving rise to photovoltages much larger than the band gap.Defects like vacancies or impurities could be of central importance[9,10,11]in this process.Moreover, the possible interesting role played by ferroelectric domain walls (DW), the boundaries between different polarization directions, are making these materials reconsidered for targeted applications[12,13,14,15,16,17]. Recently, large photovoltages observed inthin films with a high density of striped domains were attributed to band bending effects within the DWs themselves[2,4,18]. However subsequent experimental reports showed that the step in electrostatic potential in the domain wall is smaller than predicted and it was argued that the BPV effect could be at the origin of the observed behaviour[19].A related property is the finite conductivity discovered at some FE domain walls[20], also attributed to the potential step due to the abrupt change in polarization direction[18,21]. However, this picture is subject to controversy as oxygen vacancies attracted to the wallcan also explain this conduction[22,23]. Thus, it is important to clarify theexact role played by domain wallsfor the generation of photocurrents, which is the aim of the present study.

The chosen material here is $BiFeO_3$a multiferroic material combining, at room-temperature, coupled FE and antiferromagnetic orders, as well as interestingoptical properties. Its band gap is low compared to most ferroelectric perovskites, with reported values in the range of 2.6 to 3.0 eV [24,25]. This is on the edge of absorption of visible light, which has boosted theoretical and experimental investigations of this ferroelectricfor photovoltaic applications[3,26]. The samples used in the present study are high quality $BiFeO_3$ crystals grown as detailed in Ref.[27]. These are in the form of millimeter sized platelets with the short dimension along the [001] direction (in pseudocubic lattice description). The as grown samples are monodomain with the polarization along the [111] direction. 40nm thick gold electrodes weredeposited by electron beam evaporation using UV lithography in the pattern shown in Fig. 1a. Two of the electrodes are spaced by 10μm and compose the plus and minus in an electrical circuit consisting of a voltage source and a picoAmpmeter.The sample is first illuminated in a conventional wide-field microscopeby a HeNe Laser beam (λ= 632 nm)focused to illuminatethe region between electrode E1 and E2 with an estimated density around 1 kW.cm$^{-2}$. The voltage is ramped back and forth between -40V and +40V and the current variation is shown in Fig. 1.The measured I-V curves show an Ohmic behavior,from which a global bulk resistance (R ≈ 68 GΩ) can be extracted. In this geometry, theSchottky junctions established at the gold interfaces with crystalline $BiFeO_3$are almost ohmicwhich indicates that thedirection of the BFO polarization is such that it generates positive bound charges imperfectly screened by the metal electrodes[28,29,30]. As the light is turned on, a slight photoconductive effect appears at high voltage but no photovoltaic current is observed at 0 V.We attribute this phenomenon to the absence of an internal electric fieldable to effectivelyseparate the light induced electron-hole pairs. This situation is often obtained in ferroelectrics as surface screening by adsorbed species and by the electrodes can be very efficient[31]. It is also to be noted that at this wavelength, photogeneration has to come from states in the gap likely to originate from oxygen or bismuth vacancies.

In order to generateferroelectric domain walls in the measurement area,a voltage was ramped back and forthacross the electrodes, generating an electric field between +/-2.5 10$^5$V/cm. After several cycles, the I(V) curves were found to be strongly modified as shown in Fig. 1f. Besides the slight opening due to the intrinsic capacitance,a clear hysteretic behavior often associated to the presence of multidomain states[32] is observed. Interestingly, the negative voltage part of this curve is almost reversible and coming back to zero,an open circuit voltage around 1.3V associated to a photovoltaic current of 0.7 pA is obtained. When coming from large positive values, the open circuit voltage is 4.6V and the photovoltaic current 4.5 pA.Thedomain configuration was characterized (at the end of the experiments) by piezo force microscopy showing, in Fig. 1b and c,that both in-plane and out-of-plane phase contrasts are reversedroughly midway between the electrodes. Thus,the cycling procedure produced a model system with a unique DW separating two 180° domains.Under the electrodes, the polarization is found to be pointing downwards under E1 and upwards under E2 leading to the schematical configuration of Fig. 1e.

In order to studythe photo-electric propertiesof this single 180° domain wall, it is important to carry out local measurements in the vicinity ofthe DW. One possibility is to use a conducting AFM tip to locally extractthe current induced by a global illumination. Such studies have allowed to underline the special role of the tip in concentrating the electric field thereby significantly amplifying the photocurrent collection[33]. In order to better understand the topography of electron-hole pair generation, one needsrather to scan a punctual illumination source and collect the produced charges. In order to do so, we have converted an original scanning Kerr microscope into an experimental setup where a laser spot is focusedto a 600 nm spot size through a x100 (Numerical.Aperture = 0.7) objective lens which can be scanned over the sampleareabetween the electrodes.  For each position of the laser spot, two quantities are measured: the polarization rotation of the reflected light and the photocurrent collectedbetweenthe micron sized E1 and E2 electrodes. The first gives an image of the polarization domains while the second maps the local photocurrent generation. This technique provides, in a unique fashion, an interesting way to study the role of domain walls and metal/ferroelectric interfaces on photovoltaic properties.

The corresponding image of optical polar rotation of the reflected light is shown in Figure 2. As expected, no contrast is obtained between the two 180° domains. However, the dip in intensity at the DW position (where the average polarization goes to zero[34]), allows to visualize the DW during electric field sweeps. Fig. 2 showsthat the DW position can be controlled and movedhysteretically between two positions, as imaged by the optical rotation. This allows for a direct quantitative comparison of photocurrent maps for these distinct configurations under similar voltage bias (Fig. 3). In particular, one can see that for the 0V image where the DW is positioned in the middle of the gap (config. 1), the current is mainly generated when the exciting spot is close to the E1 electrodeextending over about 2μm.A clear variation of the photocurrent is observed at zero Volt for the new DW position obtained after +25V was applied (config. 2). It is now much more intense and it fills the entire space between the DW and E2(fig. 3).Thus the position of the DW leads to a fundamental difference in the photocurrent generation. It is appropriate to point out at this stage that no extra photocurrent is generated when the scanned light spot touches the DW. Instead, the PV effect is delocalized a little away from the DW, which indicates that it most likely originates from internal fieldsgenerating different conditions for the electron-hole pairs' separation. Looking deeper into the images, it can be seen that the region very close to electrode E1 plays an important role as it shows a large photocurrent intensity in config. 1. This is likely to be the hallmark of a depletion region stemming from theSchottkybarrier created by the reversed ferroelectric polarization at the Au interface. For negative bias (not shown), this depletion region grows continuously. On the opposite, for positive bias below $V_{oc}$, this region shrinks and vanishes, which is in agreement with what is expected from a Schottky contact. Above $V_{oc}$ a significant photocurrent is emitted when the region between the electrode E1 and the DW is illuminated. When the DW snaps to a new equilibrium position near V=25V (config. 2) the photocurrent clearly changes to become much more intense and located on the left part of the DW, towards the "almost" ohmic E2 electrode (fig. 3). Interestingly, the DW is quite a distorted having its left and right parts pinned, which is known to induce charging as polarization divergence is non-zero on the DW[35].

In order to confirm that the relevant quantity for the photocurrent generation is indeed the internal field, we carried out numerical simulations using a multi-physics finite elements analysis software. A 3D configuration close to that of our sample was defined with two Au/BFO interfaces. The Schottky contact is simulated taking a built-in potential value of -0.9 eV (ref [36]) and the barrier is accounted for by an interfacial effective dielectric layer with a screening length $\lambda_{eff}$ around 1nm. As shown in Fig. 4, it is possible to reproduce the observation that the Au/BFO contact goes from essentially Ohmic to Schottky when the polarization is reversed, considering a positive bound charge of about 1% of the full BFO polarization, i.e. an incomplete screening of 99%. The obtained Schottky barrier height is 1.8eV, in reasonable agreement with the measured 1.3eV. The length of the depletion region is estimated from the 0V measurement of Fig. 3 where a photocurrent is measured over a 2µm distance from the Au electrode. This is equivalent to the screening obtained assuming an oxygen vacancy concentration of $1\ 10^{-15}$ cm$^{-3}$. The inferred simulation (fig. 4a) gives a strongly localized internal field close to the Schottky contact, in the depletion region.
At positive bias, the simulated electric field is delocalized in the gap between the electrodes if the DW is considered perfectly uncharged, which does not match our photocurrent images (Fig 4b). Affecting a small negative charge to the DW (fig 4c), as argued above, allows to recover for the simulated surface electric field a shape similar to that of the photocurrent mapping (config.1). When moving the domain wall to the other position (config.2), it gets distorted in the other direction and acquires a positive charge (Fig 4d). Indeed, depending on the bending direction, the domain wall is charged negatively or positively according to the polarization divergence. This is a very reasonable hypothesis as the electric field being applied using the surface electrodes does not penetrate the full depth of our crystal (around 50µm) thus preventing the DW from moving without changing shape. The resulting simulation maps of the internal field shown in Fig. 3 (with surface charges respectively of -0.00022 C/m² and +0.00056 C/m²) reproduce quite well the photocurrent maps. In particular, for positive bias, the left side of the wall is unscreened whereas the right part is screened by the applied electric field (Fig.3 / right column). The DW is therefore becoming an active entity as it gets close to a depleted region with no charges to screen its positive polarization. The overall picture leads to a convincing scenario whereby the photocurrent is closely associated to the presence of a significant internal electric field, generated both by the Schottky contact with the Au electrodes and the bending of the DW. The latter effect is responsible for the measured impressive difference in PV images with the position and shape of the 180° DW.

In summary, our mapping of photovoltaic effects in the vicinity of a single ferroelectric 180° domain wall points to the central importance of the internal field. In line with the physical mechanisms at play in traditional photovoltaic systems, we argue that the presence of an intense local 'internal' electric field is a central prerequisite for efficient electron-hole separation. The added value of using ferroelectrics lays in the opportunity to dynamically change the internal field configuration. In particular, domain walls can be generated and positioned in order to tailor the local photovoltaic efficiency. This broadens the applicative potential of ferroelectrics for photovoltaics but also for using domain walls as electronically active entities, even though our results show that DW themselves do not exhibit specific photo-electric properties.

**Figures:**

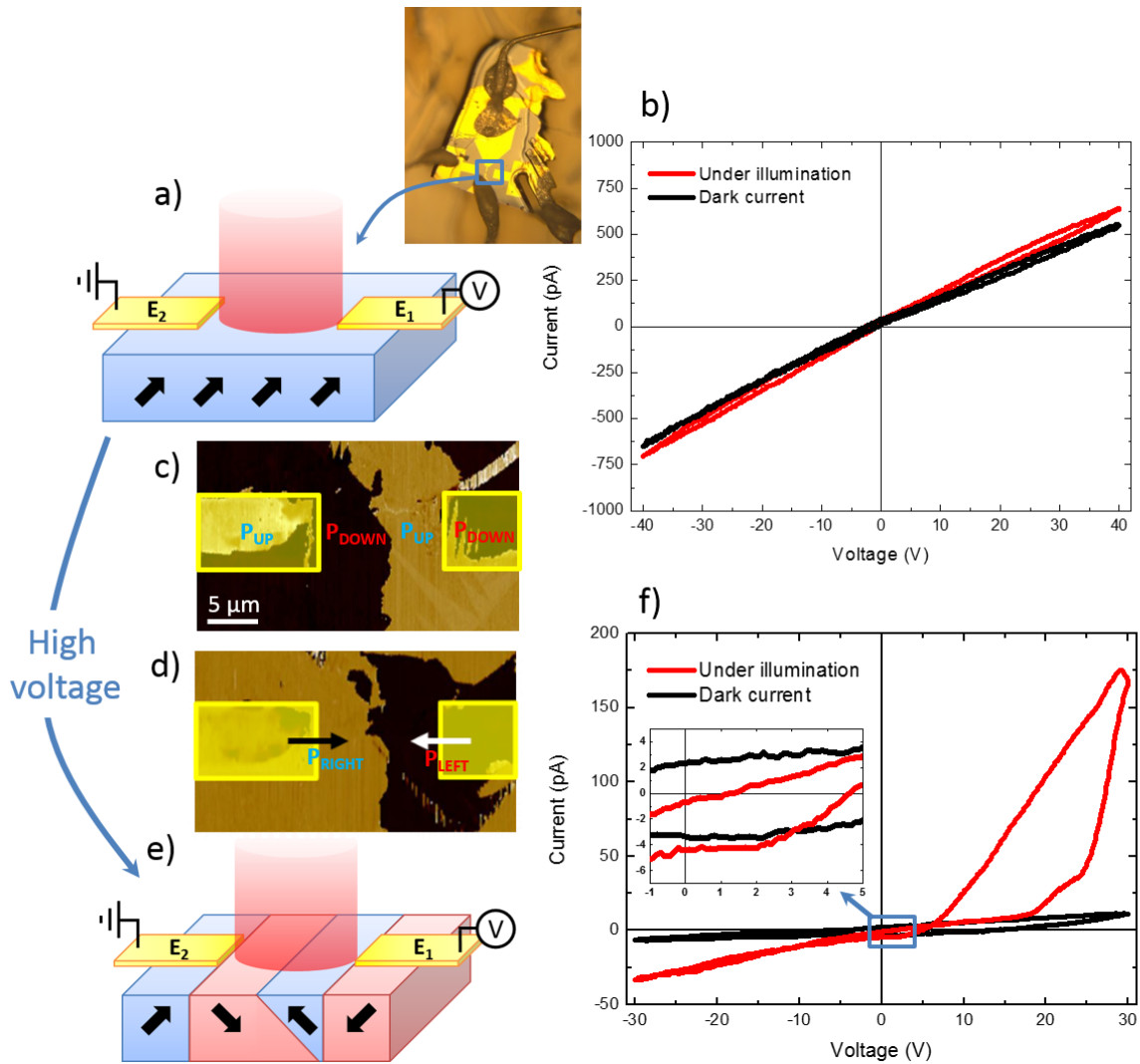

**Figure 1:** a) Single domain crystal of BiFeO$_3$ with Au electrodes deposited on its flat (001) side. b) I(V) curves measured in the dark and under illumination along with the schematics of the measurement geometry where laser light is shun in between the electrodes. A small photoconductance with negligible zero voltage photocurrent is observed. c) and d) PFM images of the sample with perpendicular (c) and planar (d) contrasts after sweeping the electric field between the electrodes, evidencing the polarization configuration schematized in e). f) The I(V) curves have dramatically changed as under illumination, a large and hysteretic current appears at positive voltages. In inset: hysteretic zero voltage currents and open circuit voltages depend on the polarization history.

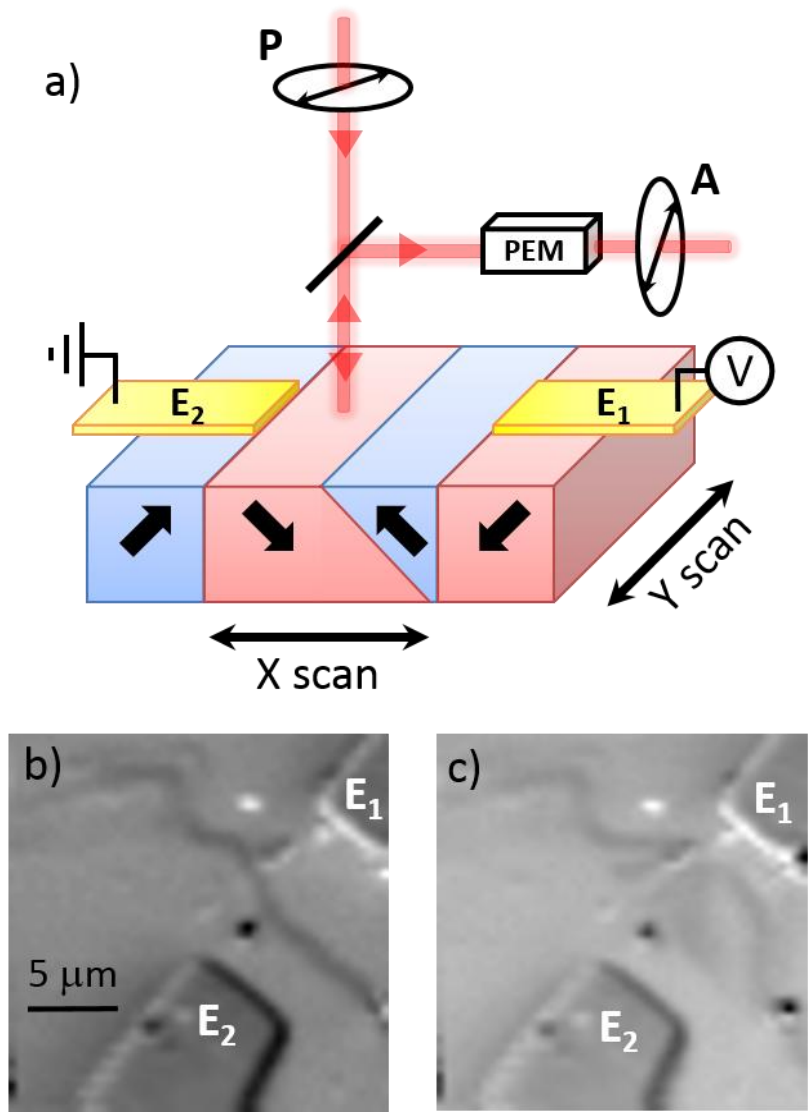

**Figure 2**: a) Schematics of the optical measurement where a focused laser spot is scanned on the sample and two quantities are recorded including the change of light polarization and the current extracted through the electrodes. The latter allows to map the photocurrent generation while the former gives the ferroelectric domain configuration as shown on the two images (b and c) where the dark line corresponds to the domain wall.

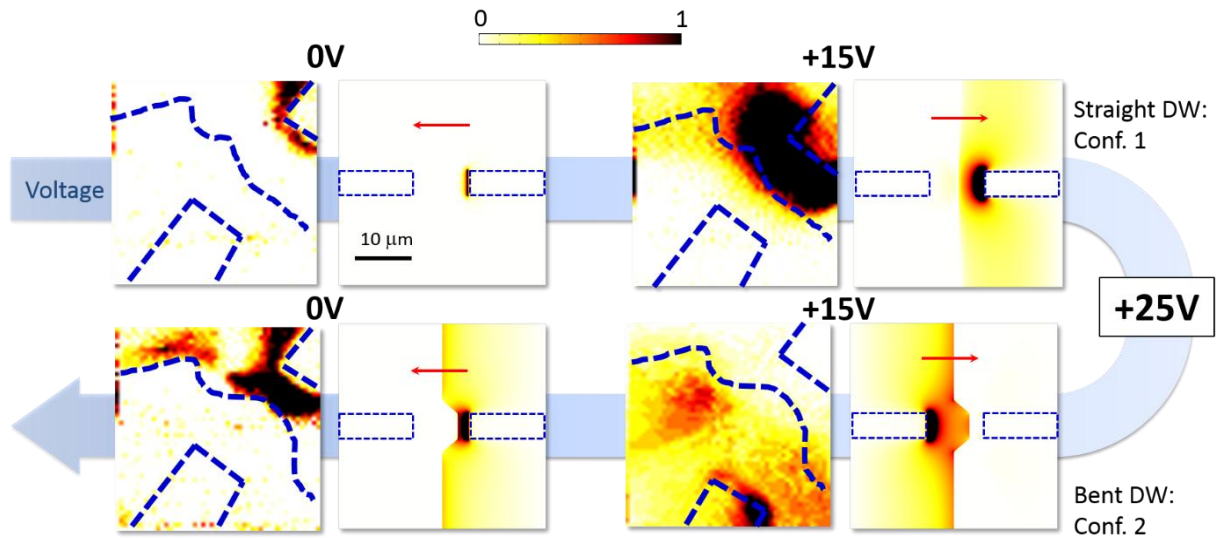

**Figure 3:** Series of photocurrent maps obtained during a (positive) voltage sweep (left column) along with simulations of the internal field (right column). The DW is, at first, as-generated by the nucleation procedure and gets subsequently displaced when applying 25V. When coming back to zero voltage, the configuration is hysteretic. The color contrast is not in absolute value but it shows clearly that the photocurrent is generated in very different regions depending on the exact position of the DW. Red arrows indicate the direction of the electrons flow. The simulations include DW charging depending on its voltage induced distortion.

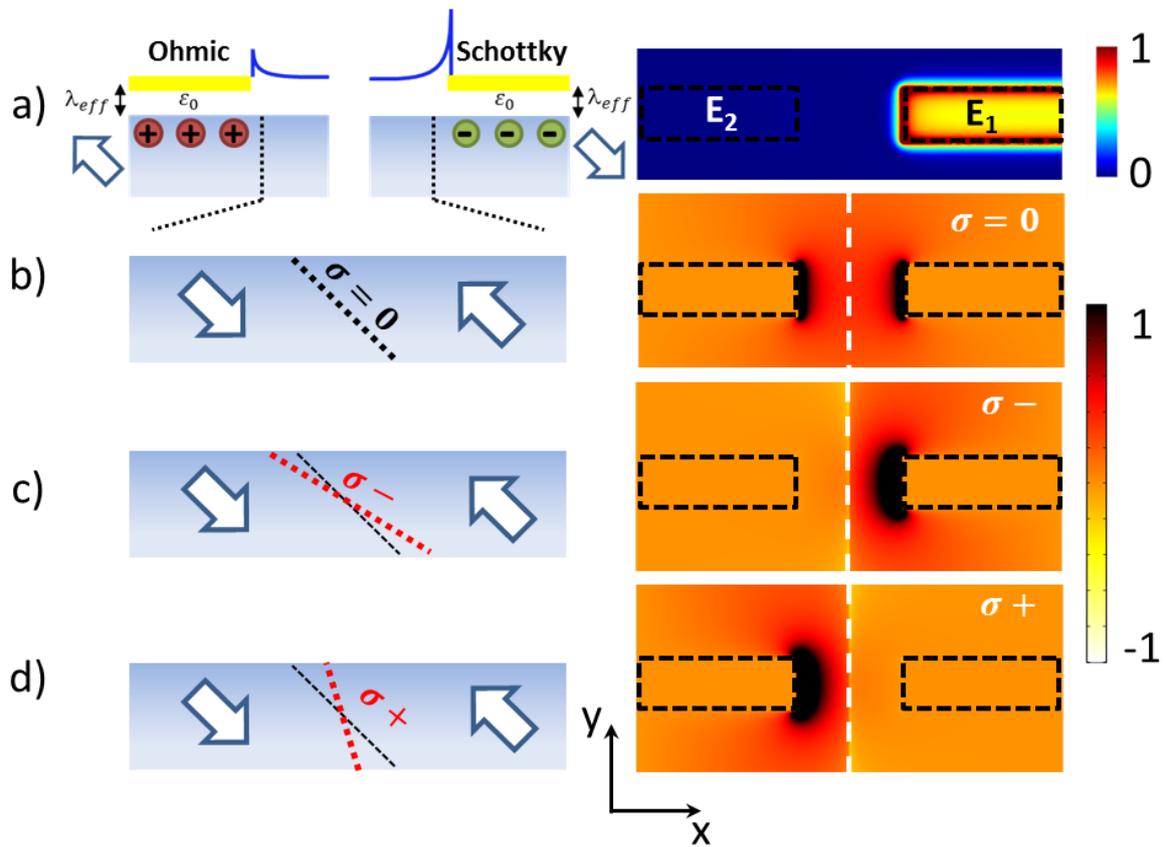

**Figure 4**: (a) Simulation of the BFO/Au contacts at 0 bias, where the direction of polarization defines either a Shottky or an (almost) Ohmic contact. (b) (c) and (d): left column shows schematic representations of the DW charge depending on the angle between P and the normal to the DW. Right column shows simulated internal field at positive bias for uncharged(b), negatively charged (c) and positively charged (d) DWs. The latter reproduces the photocurrent measurements quite well indicating that the DW is most likely slightly distorted to acquire a negative or positive charge which greatly influences the internal field. This effect, combined with the applied electric field, generates a large electric field in the region between the DW and the electrode depending on the sign of the DW surface charges. The similarity with the photocurrent mapping points to the key role played by the internal field in the PV processes.